\documentclass[preprint,12pt,authoryear,harvard]{elsarticle}
\usepackage[authoryear]{natbib}

\usepackage{amssymb}
\journal{JMPS}

\begin{document}
\newcommand{\etal}{{\em et al.}}{}
\newcommand{\fig}[1]{Fig.~\ref{#1}}
\newcommand{\tab}[1]{{Table ~(\ref{#1})}}
\newcommand{\eqn}[1]{{Eq.~(\ref{#1})}}
\newcommand{\AAA}{\AA\,}{}    % angstrom
\newcommand{\eVA}{$\rm{eV/\rm{\AA}}$\,}{} % unit of force
\newcommand{\ww}{1.0}
%%\renewcommand{\citet}[1]{\citeauthor{#1} (\citeyear{#1})}
%%\setcitestyle{authoryear,round,comma,aysep={:},yysep={,},notesep={,}}
\begin{frontmatter}



\title{A comparative study of fracture in Al: quantum mechanical vs. empirical atomistic description}


\author{Qing Peng and Gang Lu}

\address{ Department of Physics and Astronomy, California State
University Northridge, \\Northridge, CA, USA}

\begin{abstract}
A comparative study of fracture in Al is carried out by using
quantum mechanical and empirical atomistic description of atomic
interaction at crack tip. The former is accomplished with the
density functional theory (DFT) based Quasicontinuum method (QCDFT)
and the latter with the original Quasicontinuum method (EAM-QC).
Aside from quantitative differences, the two descriptions also yield
qualitatively distinctive fracture behavior. While EAM-QC predicts a
straight crack front and a micro-twinning at the crack tip, QCDFT
finds a more rounded crack profile and the absence of twinning.
Although many dislocations are emitted from the crack tip in EAM-QC,
they all glide on a single slip plane. In contrast, only two
dislocations are nucleated under the maximum load applied in QCDFT,
and they glide on two adjacent slip planes. The electron charge
density develops ``sharp corners" at the crack tip in EAM-QC, while
it is smoother in QCDFT. The physics underlying these differences is
discussed.
\end{abstract}


\end{frontmatter}





\section{Introduction}
Understanding fracture behavior in materials is a challenging
undertaking. Despite nearly a century of study, several important
issues remain unsolved. For example, there is little fundamental
understanding of brittle to ductile transition as a function of
temperature in many materials; there is still no definitive
explanation of how fracture stress is transmitted through plastic
zones at crack tips; and there is no complete understanding of the
disagreement between theory and experiment regarding the limiting
speed of crack propagation. These difficulties to a great extent
stem from the fact that fracture phenomena are governed by processes
occurring over a wide range of length and time scales; these
processes are all connected and all contribute to the total fracture
energy \citep{needleman}.

As emphasized by Van de Giessen and Needleman, although the
atomistic interaction at a crack tip may only account for a small
fraction of the total fracture energy, it can be a controlling
factor \citep{needleman} - after all, all fractures take place by
breaking atomic bonds. Critical atomistic information, such as
surface energy, stacking fault energy, dislocation
nucleation/propagation energy and twinning formation energy, etc,
has been known to play central roles in fracture. In fact, some of
these quantities are at the heart of fracture mechanics, including
the Griffith's criterion for brittle fracture \citep{Griffith1},
Rice's criterion \citep{Rice1992} for crack tip blunting and more
recently a criterion for twinning at crack tip \citep{Tadmor}, to
name a few.

Because of the inherent multiscale nature of fracture - the process
at each scale depends strongly on what happens at the other scales,
the modeling and simulation of fracture calls for {\it concurrent}
multiscale approaches \citep{review}. One of the first concurrent
multiscale modeling of fracture was based on Macroscopic Atomistic
{\it ab initio} Dynamics (MAAD) method for Silicon \citep{maad}.
MAAD couples a quantum mechanical description of atoms at crack tip,
to an empirical (or classical) atomistic description of atoms at a
short distance away from the crack tip, and to the continuum
finite-element description of the rest of the system. Since MAAD,
several other concurrent multiscale methods have been developed, all
involving some level of quantum mechanical modeling at the crack tip
\citep{Csanyi, Bernstein,Ogata}. All these methods were
developed/applied for Si owning to the following technical reasons:
(1) large-scale electronic structure methods such as linear-scaling
algorithms are only applicable to covalently-bonded semiconductors
like Si; general approaches for metals still remain elusive; (2)
satisfactory QM/MM coupling schemes for metals were less well
developed until recently \citep{Bernsteinreview, Yi, Xu}. On the
other hand, concurrent multiscale approaches that do not involves
quantum mechanics are readily available
\citep{miller,Kohlhoff1991,Buehler06}. Among them, Quasicontinuum
(QC) method is particularly promising and it has been widely applied
to many materials problems, including fracture in metals
\citep{QCreview}. QC strives to achieve a ``seamless" coupling
between atomistic and continuum descriptions and allows quantum
mechanical interactions incorporated in a systematical manner. For
example, although the original QC was based on classical atomic
interactions, significant progress has been made recently to
incorporate quantum mechanical interactions in the local QC region
\citep{LQC2006}, nonlocal QC region \citep{Luqcdft} and entire QC
system \citep{peng08}. The coarse-graining strategy of QC has also
been explored to perform large-scale electronic structure
calculations \citep{Gavini}.

Despite the impressive advance in multiscale methodology
development, a crucial question remains unanswered. Although it is
clear that an atomistic description at a crack tip is indispensable
for many purposes, it is not well established whether a quantum
mechanical description at the crack tip is truly necessary. This is
a poignant point given the continuing improvement of empirical
potentials and the still heavy costs for quantum simulations. It is
to address this question that motivates the present study. As a
first look at the problem, we focus on crack tip plasticity in Al
and compare results received from a quantum mechanical description
vs. an empirical atomistic description at the crack tip, both in the
framework of QC. Since the atomistic resolution is only necessary
near the crack tip while the linear elastic fracture mechanics
boundary conditions can be applied in the far field, QC is well
poised for such fracture simulations. In addition, QC simulations
involve quasi-static energy minimization, thus the unrealistically
high strain rates common to molecular dynamics simulations are
avoided. Unfortunately, as a result thermally activated processes
are precluded in QC. Al is chosen in this study because it is
relatively inexpensive for density functional theory (DFT)
calculations and an excellent embedded-atom-method (EAM) empirical
potential exists for Al \citep{eam}. The goal of this work is to
examine how and why the results received in the empirical
simulations differ from those in quantum mechanical DFT simulations
at the crack tip. To this end, we employ so-called QCDFT method in
which the nonlocal atoms at the crack tip are treated with DFT. The
QCDFT results are compared to those obtained from the original QC
method in which all nonlocal atoms are treated with EAM empirical
potential.

The structure of the paper is as follows. The methodology is
introduced in Section 2 for both QC and QCDFT methods. A
semi-infinite crack under mode I loading is set up in Section 3.1.
The computational parameters are described in Section 3.2 and the
loading procedure for the crack is summarized in Section 3.3. The
simulation results and analysis are presented in Section 4 and
discussions are given in Section 5. Finally we conclude in Section
6.

\section{Methodology}

The QC method \citep{qc1,qc2} is a multiscale approach
\citep{review} that combines atomistic models with continuum
theories, and thus offers an advantage over conventional atomistic
simulations in terms of computational efficiency. The idea
underlying the QC method is that atomistic processes of interest
often occur in very small spatial domains (e.g., crack tip) while
the vast majority of atoms in the material behave according to
well-established continuum theories. To exploit this fact, the QC
method retains atomic resolution only where necessary and coarsens
to a continuum finite element description elsewhere. This is
achieved by replacing the full set of $N$ atoms with a small subset
of $N_r$ ``representative atoms'' or {\it repatoms} ($N_r\ll N$)
that approximate the total energy through appropriate weighting. The
energies of individual repatoms are computed in two different ways
depending on the deformation in their immediate vicinity. Atoms
experiencing large deformation gradients on an atomic-scale are
computed in the same way as in a standard fully-atomistic method. In
QC these atoms are called {\em nonlocal} atoms. In contrast, the
energies of atoms experiencing a smooth deformation field on the
atomic scale are computed based on the deformation gradient in their
vicinity as befitting a continuum model. These atoms are called {\em
local} atoms. The total energy $E_{\rm tot}$ (which for a classical
system can be written as $E_{\rm tot}=\sum_{i=1}^N E_i$, with $E_i$
the energy of atom $i$) is approximated as
\begin{equation}
E_{\rm {tot}}^{\rm {QC}}=\sum_{i=1}^{N^{{\rm nl}}}E_i(\{{\bf q}\})
+\sum_{j=1}^{N^{{\rm loc}}}n_jE_j^{{\rm loc}}(\{{\bf F}\}).
 \end{equation}
The total energy has been divided into two parts: an atomistic
region of $N^{\rm nl}$ nonlocal atoms and a continuum region of
$N^{{\rm loc}}$ local atoms ($N^{{\rm nl}}+N^{{\rm loc}}=N^r$).

The original formulation of QC was limited to classical potentials
for describing interactions between atoms. However, since many
materials properties depend crucially on the behavior of electrons,
such as bond breaking/forming at crack tips or defect cores,
chemical reactions with impurities, surface reactions and
reconstructions, electron excitation and magnetism, etc, it is
desirable to incorporate appropriate quantum mechanical descriptions
into the QC formalism. QCDFT is one strategy to fill this role. In
specific, QCDFT combines the coarse graining idea of QC and the
coupling strategy of the quantum mechanics/molecular mechanics
(QM/MM) method. This method can capture the electronic structure at
the crack tip within the accuracy of DFT and at the same time reach
the length-scale that is relevant to
experiments\citep{Luqcdft,peng08}.

The original QC formulation assumes that the total energy can be
written as a sum over individual atom energies. This condition is
not satisfied by quantum mechanical models. To address this
limitation, in the present QCDFT approach the nonlocal region is
treated by an EAM-based QM/MM coupling approach \citep{Luqcdft, Yi}:
the Kokn-Sham density functional theory (KS-DFT) is coupled to EAM
with the interaction energy calculated also by EAM. The local
region, on the other hand, is dealt with by EAM, which is the same
energy formulation used in the MM part of the nonlocal region. This
makes the passage from the atomistic to continuum seamless since the
same underlying material description is used in both. This
description enables the model to adapt automatically to changing
circumstances (e.g. the nucleation of new defects or the migration
of existing defects). The adaptability is one of its main strengths
of QC and QCDFT, which is missing in many other multiscale methods.

More specifically, in the present QCDFT approach the material of
interest is partitioned into three distinct types of domains: (1) a
nonlocal quantum mechanical DFT region (region I); (2) a nonlocal
classical region where classical EAM potentials are used (region
II); and (3) a local region (region III) that employs the same EAM
potentials as region II. The coupling between regions II and III is
achieved via the QC formulation, while the coupling between regions
I and II is accomplished by the QM/MM scheme \citep{Choly2005, Yi}.
The total energy of the QCDFT system is then \citep{Luqcdft}
\begin{equation}
\begin{array}{c}
\label{energyqcdft} E_{\rm {tot}}^{\rm {QCDFT}} = E^{{\rm nl}}[{\rm
I+II}]
+\sum_{j=1}^{N^{{\rm loc}}}n_jE_j^{{\rm loc}}(\{{\bf F}\}) \\
=E_{\rm{DFT}}[{\rm I}]-E_{\rm{EAM}}[{\rm I}] +E_{\rm{EAM}}[{\rm
I+II}]+ \sum_{j=1}^{N^{{\rm loc}}}n_jE_j^{{\rm loc}}(\{{\bf F}\}),
\end{array}
\end{equation}
where $E^{{\rm nl}}[{\rm I+II}]$ is the total energy of the nonlocal
region (I and II combined with the assumption that region I is
embedded within region II), $E_{\rm{DFT}}$[I] is the energy of
region I in the absence of region II computed with DFT,
$E_{\rm{EAM}}$[II] is the energy of region II in the absence of
region I computed with EAM, and $E_{\rm{EAM}}[{\rm I+II}]$ is the
energy of the nonlocal region computed with EAM.

Other types of combination with quantum mechanical and classical
atomistic methods may also be implemented in QCDFT. The great
advantage of the present implementation is its simplicity; it
demands nothing beyond what is required for a DFT calculation and an
EAM-QC calculation. Another important practical advantage of QCDFT
method is that, if region I contains many different atomic species
while region II contains only one atom type, there is no need to
develop reliable EAM potentials that can describe each species and
their interactions. This is because if the various species of atoms
are well within region I, the energy contributions of these atoms
are canceled out in the total energy calculation. This advantage
renders the method particularly useful in dealing with impurities,
which is an exceedingly difficult task for empirical potential
simulations.

The equilibrium structure of the system is obtained by minimizing
the total energy in Eq.~\ref{energyqcdft} with respect to all
degrees of freedom. Because the time required to evaluate
$E_{\rm{DFT}}$[I] is considerably more than that required for
computation of the other EAM terms in $E_{\rm {tot}}^{\rm {QCDFT}}$,
an alternate relaxation scheme turns out to be useful. The total
system can be relaxed by using conjugate gradient approach on the
DFT atoms alone, while fully relaxing the EAM atoms in region II and
the displacement field in region III at each step. An auxiliary
energy function can be defined as
\begin{equation}
\label{relax} E'[\{{\bf q}^{\rm I}\}] \equiv \min_{\{{\bf q}^{\rm
II}\},\{{\bf q}^{\rm III}\}} E_{\rm tot}^{\rm QCDFT}[\{{\bf q}\}],
\end{equation}
which allows for the following relaxation scheme: (i) minimize
$E_{\rm tot}^{\rm QCDFT}$ with respect to the atoms in regions II
($\{{\bf q}^{\rm II}\}$) and the atoms in region III ($\{{\bf
q}^{\rm III}\}$), while holding the atoms in region I fixed; (ii)
calculate $E_{\rm tot}^{\rm QCDFT}[\{{\bf q}\}]$, and the forces on
the region I atoms; (iii) perform one step of conjugate gradient
minimization of $E'$; (iv) repeat until the system is relaxed. In
this manner, the number of DFT calculations performed is greatly
reduced, albeit at the expense of more EAM and local QC
calculations. A number of tests have shown that the total number of
DFT energy calculations for the relaxation of an entire system is
about the same as that required for DFT relaxation of region I
alone.

\section{Computational details}
\subsection{Model setup}
\begin{figure}[htp]
\centering
\includegraphics[width=\ww\textwidth]{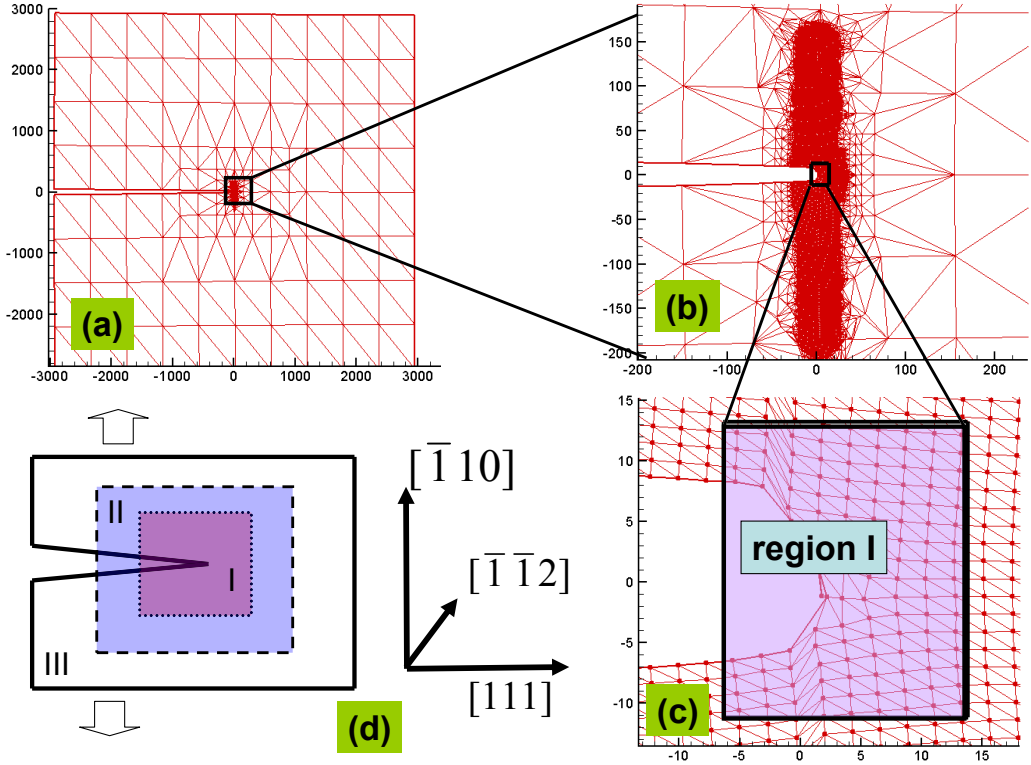}
 \caption{\label{fig:conf} (Color online) (a) The overview of the entire crack system with finite-element
 mesh; (b) A blown-up view of (a) showing the nonlocal region; (c) region I box
 and atomic structure at the crack tip; (d) schematic partition of the system into region I, II and III.
 The $x$, $y$ and $z$ axis is along [111],[$\bar{1}$10], and [$\bar{1}\bar{1}2$],
respectively.  All lengths are in \AA. }
\end{figure}

A semi-infinite crack in a single Al crystal is studied by both QC
and QCDFT for comparisons. The crack is made by removing two layers
of atoms with $x<0$ and $y=0 \,\&\, 1.41$ \AA. The crack plane is
({\rm $\bar{1}$10}) and in this orientation, (111) plane is the only
active slip plane for dislocations emitted from the crack tip; all
other $\{111\}$-type planes lie obliquely to the crack plane and are
thus precluded by the imposed plane strain conditions
\citep{Tadmor}. This configuration was used previously in a MD study
by Hoagland {\it et al.} \citep{Hoagland} and an EAM-QC study by Hai
{\it et al.} \citep{Tadmor_acta} although the initial crack opening
and the EAM potential used are different in these studies. The
initial crack opening in the present work is determined based on two
competing considerations: (1) it cannot be too narrow otherwise the
crack will close and/or large number of loading steps is required to
observe the onset of plasticity; (2) it cannot be too wide otherwise
the DFT region is too large to render calculations feasible. Of
course, the DFT region has to be large enough to capture the crucial
plasticity events at the tip. The crack is subject to mode I loading
along $y$ direction shown schematically in \fig{fig:conf}(d).

The dimension of the system is $6000 \times 6000 \times 4.887$ ${\rm
\AA}^3$ along the $x$, $y$, $z$ directions respectively. The system
is periodic in z-direction, and has Dirichlet boundary conditions in
the other two directions. The system contains over 11 million Al
atoms - a size that is well beyond the reach of any full-blown
quantum calculation. The schematic overview of the system is shown
in panel (a) of \fig{fig:conf} with finite element meshes. Panel (b)
is a zoomed-in view of the nonlocal region and panel (c) displays
the atomic detail of the crack tip.

\subsection{Computational parameters}
Two comparative calculations are carried out for the crack: EAM-QC
vs. QCDFT for Al. EAM-QC is the original QC with EAM potential for
atomic interactions. In this work, the EAM potential used is
rescaled from the original ``force-matching" EAM \citep{eam}
potential so that it matches precisely the value of the lattice
constant and bulk modulus of Al from the DFT calculations
\citep{Choly2005}. Although the re-scaling changes very little to
the original potential, it eliminates lattice parameter mismatch at
the QM/MM interface, and thus reduce QM/MM coupling errors.

DFT calculations are carried out with the Vienna Ab-initio
Simulation Package (VASP)
\citep{Kresse1993,Kresse1994,Kresse1996b,Kresse1996a} which is based
on Kohn-Shem Density Functional Theory (KS-DFT) with the local
density approximation and ultrasoft pseudopotentials. A plane-wave
cutoff of 129 eV is used in the calculations and the $k$-points are
sampled according to the Monkhorst-Pack method \citep{MP} with a $1
\times 1 \times 9$ mesh in the Brillouin zone. There are 134 DFT
atoms in region I.

\subsection{Loading procedure}
The simulations are performed quasi-statically with displacement
boundary conditions where the displacement is prescribed as a
function of intended stress intensity factor (SIF) at each loading
step. For EAM-QC, the loading procedure is outlined as following: \\
(1) For a small initial stress intensity factor $K_{\rm{I}}$, the
anisotropic Linear Elastic Fracture Mechanics (LEFM) solution
\citep{LEFM} $\rm{\bf{u}}_{LEFM}(\rm{\bf{X}},K_{\rm{I}})$ is
obtained. Each atom is displaced according to
$\rm{\bf{u}}(\rm{\bf{X}})=\rm{\bf{u}}_{LEFM}(\rm{\bf{X}},K_{\rm{I}})$,
where $\rm{\bf{u}}$ is the displacement field and $\rm{\bf{X}}$ is the position of nodes in the model.  \\
(2) The displacement at the model boundaries is fixed except for
the crack surfaces; the positions for all repatoms are obtained by energy minimization as explained before. \\
(3) The finite-element mesh, the status of repatoms (local vs.
nonlocal) and the neighbor list are updated. \\
(4) The SIF is increased by a small amount $\Delta K_{\rm{I}}$ as
$K_{\rm{I}}=K_{\rm{I}}+\Delta K_{\rm{I}}$ and repeat from step (1) until the intended SIF is achieved. \\
In this study the increment $\Delta K_{\rm{I}}=0.001$ ${\rm
eV/\AA}^{2.5}$ is used. The loading procedure adopted here follows
closely that of reference \citep{Tadmor_acta}.

Because DFT calculations are much more expensive than EAM, we use
EAM-QC to load the crack until an incipient plasticity is about to
take place, at which point QCDFT is switched on. In other words, a
QCDFT relaxation starts from a configuration that is obtained by
EAM-QC for $K_{\rm{I}}$. QCDFT then increases the SIF by $\Delta
K_{\rm{I}}$. This is a reasonable approximation because EAM is known
to give accurate results for deformations prior the onset of crack
tip plasticity or the appearance of lattice defects. As will be
shown later, since the critical load for the onset of plasticity
from QCDFT is smaller than that from EAM-QC, the present loading
strategy does not run the risk of missing incipient plasticity of
QCDFT.

\section{Results and Analysis}

\subsection{EAM-QC calculation of Al}
\begin{figure}[htp]
\centering
\includegraphics[width=0.6\textwidth]{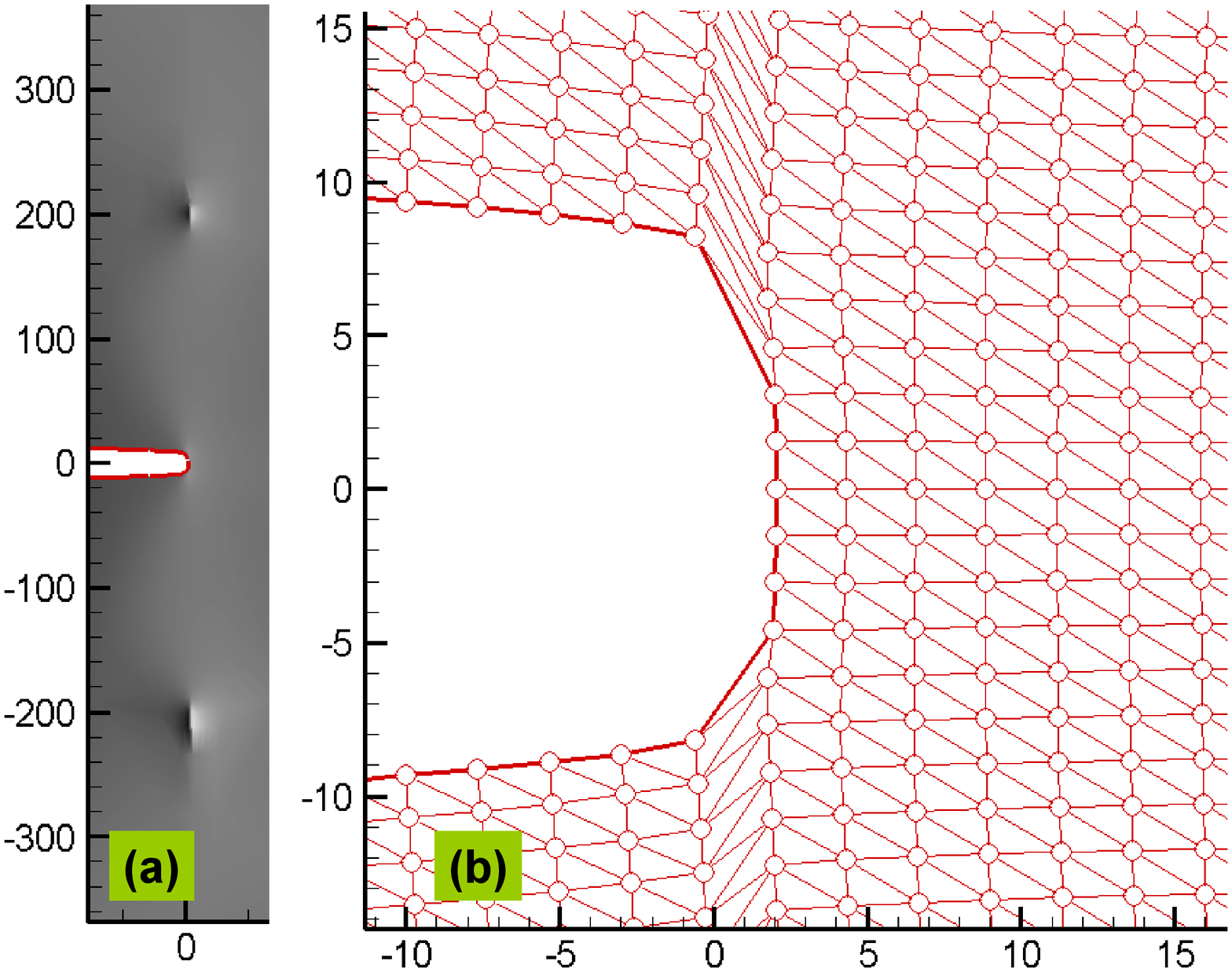}
\label{fig:originalQC} \caption{EAM-QC results at SIF $K_{\rm{I}} =
0.144 \,{\rm eV/\AA^{2.5}}$. (a)The out-of-plane displacement $U_z$;
the displacement contours range from -0.5 \AA (darkest) to 0.5 \AA
(lightest). (b)The zoomed-in view of the crack tip atomic structure
and finite-element mesh. The open circle represents the atomic
position. All distances are in \AA.}
\end{figure}
To establish the validity of present EAM-QC method, we first perform
EAM-QC calculations for the same crack studied in reference
\citep{Tadmor_acta}. We use the same EAM potential and the same
initial crack opening as that in \citep{Tadmor_acta}. The results
are presented in \fig{fig:originalQC} where two edge dislocations
are emitted from the crack tip and they glide at the same slip plane
in a symmetrical manner. The critical SIF is 0.144 ${\rm
eV/\AA^{2.5}}$. All these results are identical to those found in
\citep{Tadmor_acta} and thus validate the present EAM-QC method.

\begin{figure}[htp]
\centering
\includegraphics[width=\ww\textwidth]{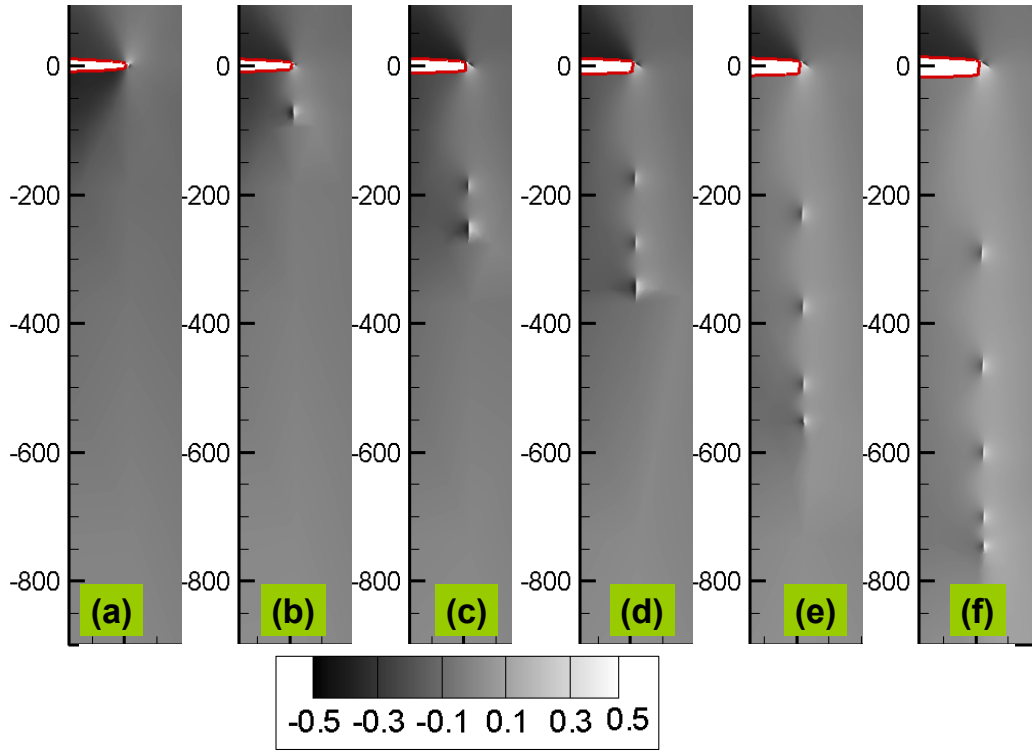}
 \caption{\label{fig:eam} The out-of-plane displacement $U_z$
obtained from the EAM-QC calculations at SIF $K_{\rm I}$ of (a)
0.179, (b) 0.180, (c) 0.184, (d) 0.186, (e) 0.191, and (f) 0.198
${\rm eV/\AA}^{2.5}$ respectively. The displacement contours range
from -0.5 \AA (darkest) to 0.5 \AA (lightest). The crack surfaces
are represented by red curves. All distances are in \AA.}
\end{figure}

Next, we apply EAM-QC to the crack of interest. The crack is loaded
quasi-statically and no crack tip plasticity is observed until the
SIF reaches $K_{\rm{I}} =0.180$ ${\rm eV/\AA}^{2.5}$. By comparison,
the critical SIF for pure brittle cleavage computed from the
Griffith criterion for this orientation is $K_{\rm Ic} = 0.205
\,{\rm eV/\AA}^{2.5}$. In \fig{fig:eam}, we present the out-of-plane
displacement $U_z$ as a function of applied $K_{\rm{I}}$ values. For
$K_{\rm{I}} =0.179$ ${\rm eV/\AA}^{2.5}$, although no dislocation is
observed, significant deformation at the crack tip is clearly
visible (panel a). At $K_{\rm{I}}=0.180$ ${\rm eV/\AA}^{2.5}$, the
first dislocation is nucleated and subsequently moves away from the
crack tip. The dissociated $1/2[1\bar{1}0]$ edge dislocation is
stabilized at about 70 \AA\, below the crack on a (111) plane (panel
b). The contour shading of \fig{fig:eam} corresponds to the
magnitude of $U_z$, whose non-zero values indicate that the edge
dislocation is dissociated into partials. Interestingly, at the same
time, a micro-twin is also nucleated from the crack tip
(\fig{fig:twin}a). The micro-twn is two layers in both length and
width, the twinning direction is [714] and the twinning plane is
($1\bar{3}\bar{1}$). As $K_{\rm{I}}$ value is increased, more
dislocations are emitted on the (111) plane and {\it at the same
time}, the micro-twin grows in length but not in width. More
specifically, as $K_{\rm{I}}$ increases to 0.184, 0.186, 0.191,
0.198 ${\rm eV/\AA}^{2.5}$, two, three, four and five dislocations
are emitted from the crack tip and they glide on the same (111)
plane, as shown in the panel of (c), (d), (e) and (f) of
\fig{fig:eam} respectively. Correspondingly, the micro-twin grows to
three, four, five and six layers in length, respectively. The
micro-twin structures of two, three, five and six layers in length
are shown in the panel of (a), (b), (c), (d) of \fig{fig:twin}
respectively. The width of the micro-twin is not increasing perhaps
due to unfavorable stacking fault energy along the usual twinning
plane. The reason that the twinning was not observed in Hai {\it et
al.} \citep{Tadmor_acta} is probably due to the the different
initial crack openings rather than the different EAM potentials
used. We have done additional calculations for the present crack
opening (2 layers) with the same EAM potential used in reference
\citep{Tadmor_acta} and found a similar twinning at the crack tip.
\begin{figure}[htp]
\centering
\includegraphics[width=0.6\textwidth]{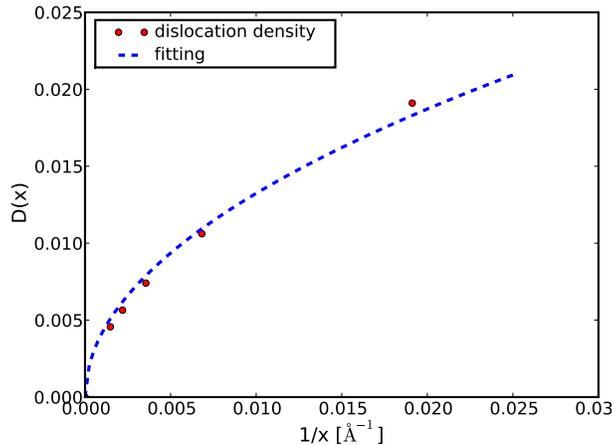}
 \caption{\label{fig:disden} The dislocation distribution function D(x) as
a function of the inverse of the distance x. The dashed curve is a
fit to the filled circles.}
\end{figure}
The emitted dislocations share the following characteristics: (1)
They are all edge dislocations of ${\rm 1/2[1\bar{1}0]}$ Burgers
vector and dissociated into Shockly partials with a separation
distance of 16 \AA. This result agrees with the previous study of
Hai {\it et al.} \citep{Tadmor_acta} with the similar EAM potential.
But this value is too large compared to the experimental splitting
distance of 5.5 \AA , measured by Mills and Stadelmann\citep{Mills}.
The discrepancy may be attributed to still too low stacking fault
energy of the EAM potential. (2) They are all on the same $\{111\}$
slip plane whose position is $ x = a/2$, where is $a$ is (111)
inter-plane distance. The active slip plane is slightly ahead of the
crack front position at $x = 0$. The emitted dislocations move away
from the crack tip and form a pile-up against the local/nonlocal QC
interface. For $K_{\rm{I}}=0.198 \,{\rm eV/\AA}^{2.5}$, the
dislocation density $D(x)$ defined as the number of dislocations per
unit distance along pileup line, is found to be the square root of
the inverse of the distance \citep{disdensity}. The distance $x$
refers to the distance between the dislocation center and the
local/nonlocal interface. The fitted curve (dashed line) in
\fig{fig:disden} qualitatively agrees well with the elastic theory
where the fitting function is given as $D(x)=0.13236 (1/x)^{1/2}$.
\begin{figure}[htp]
\centering
\includegraphics[width=\ww\textwidth]{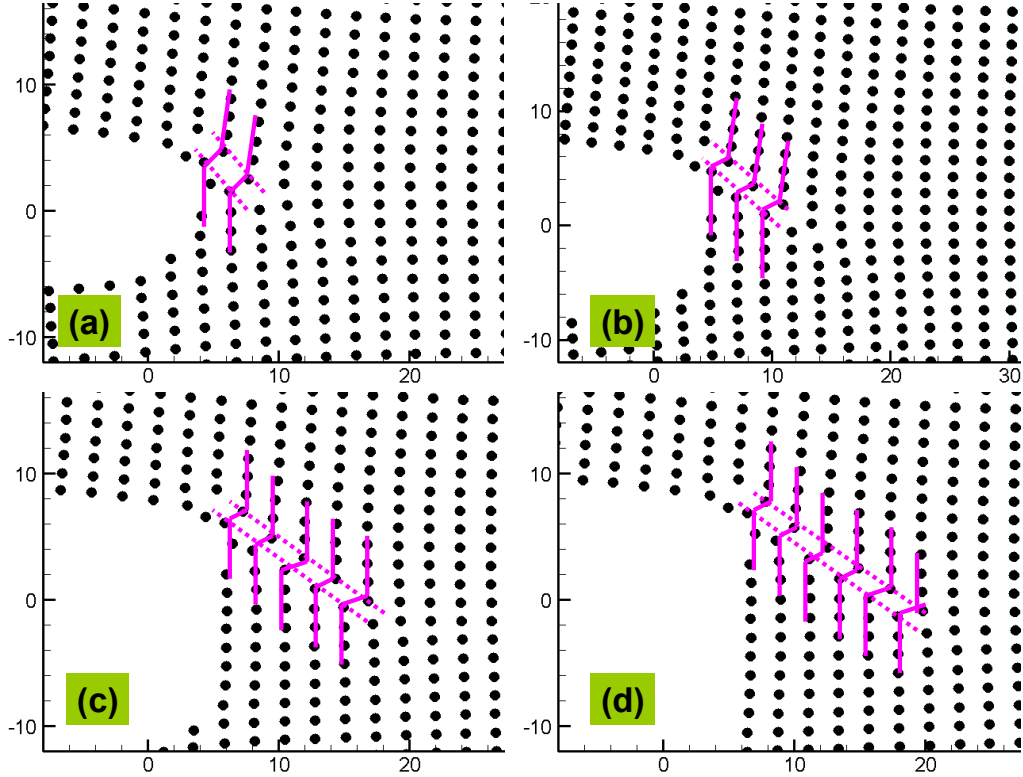}
 \caption{\label{fig:twin} Atomic structure at the crack tip from EAM-QC.
Some of atomic planes are highlighted in pink to indicate the twin.
Dashed lines represent the two-layer twin. (a) $K_{\rm I}$ = 0.180,
 (b) $K_{\rm I}$ = 0.184, (c) $K_{\rm I}$ = 0.191, (d) $K_{\rm I}$ = 0.198 ${\rm eV/\AA}^{2.5}$. All distances are in \AA.}
\end{figure}

\subsection{QCDFT calculation of Al}
The crack is first loaded by EAM-QC until $K_{\rm{I}}$ reaches 0.169
${\rm eV/\AA}^{2.5}$ at which point QCDFT is started. This critical
loading is determined by trial and error; we launch a number of
QCDFT calculations at different $K_{\rm{I}}$ from previously relaxed
EAM-QC structures and examine whether any crack tip plasticity is
taking palce. The smallest $K_{\rm{I}}$ value that results in
incipient plasticity is the critical loading. In comparison, the
critical SIF for pure brittle cleavage computed from the Griffith
criterion is $K_{\rm Ic} = 0.267 \,{\rm eV/\AA}^{2.5}$.

\begin{figure}[htp]
\centering
\includegraphics[width=\ww\textwidth]{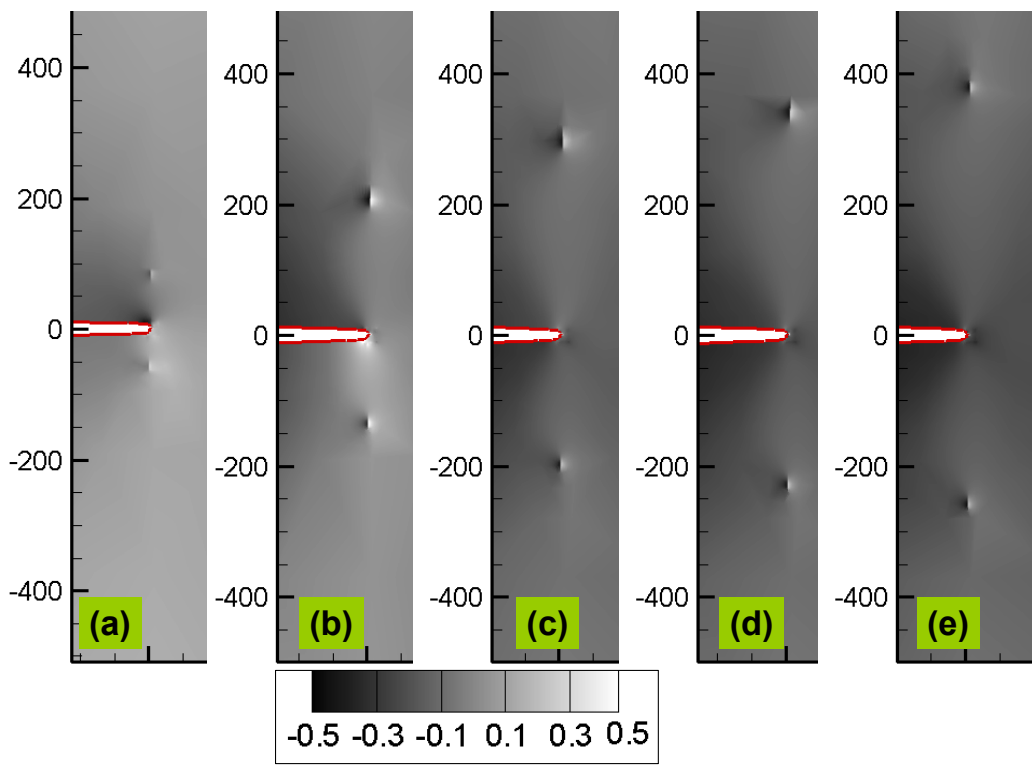}
\caption{\label{fig:QCDFT} The out-of-plane displacement $U_z$
obtained from QCDFT calculations at SIF $K_{\rm I}$ of (a) 0.169,
(b) 0.170, (c) 0.174, (d) 0.175 (e) 0.178 ${\rm eV/\AA}^{2.5}$
respectively. The displacement contours range from -0.5 \AA
(darkest) to 0.5 \AA (lightest). All distances are in \AA. }
 \end{figure}

The crack tip behavior of QCDFT is much different. At
$K_{\rm{I}}=0.169$ ${\rm eV/\AA}^{2.5}$, two dissociated edge
dislocations are nucleated - one above the crack plane and one below
it. In contrast to EAM-QC, the two dislocations glide at two {\it
adjacent} (111) slip planes, as shown in Fig. 7(d). The positions of
the two slip planes are at $x = - a/2 $ and $x = -3a/2$
respectively; they are slightly behind the crack front position. The
separation distance of the two Schockly partials is 16 \AA, the same
as that in EAM-QC calculation. This result is consistent with the
fact that the dislocation cores are outside the DFT box where atomic
interaction is determined by the rescaled EAM potential.

As the load is increased, the two emitted dislocations are driven
further away from the crack plane, however, no more dislocation is
nucleated within the maximum load that we have explored in this
study. For the largest $K_{\rm{I}} = 0.178 \,{\rm eV/\AA}^{2.5}$,
there are only two emitted dislocations gliding at 380 \AA\ (up) and
258 \AA\ (down) away from the crack plane. Due to the computational
cost of QCDFT, we did not pursue more calculations for even larger
loadings. However, it is evident from the present study that the
crack tip plasticity observed in QCDFT is {\it qualitatively
different} from that in EAM-QC. The differences are more striking
given the fact that the QCDFT results are continued relaxation of
EAM-QC configurations.

\section{Discussion}

\begin{table}[ht]
\begin{minipage}[b]{1.0\linewidth}
 \caption{Relevant quantities calculated by VASP and EAM for bulk Al; the corresponding experimental values are
extrapolated to T=0 K.} % title of Table
\begin{tabular}{|c|c|c|c|c|c|c|c|c|}
    \hline
 &$a_0$&$\gamma_{111}$&$\gamma_{110}$&$\gamma_{210}$&$\gamma_{sf}$&$\gamma_{us}$&$\gamma_{ut}$&$\tau_t$ \\

 &(\AA)&${(\rm J/m}^2)$&$({\rm
 J/m}^2)$&$({\rm J/m}^2)$&$({\rm
 J/m}^2)$&$({\rm J/m}^2)$&$({\rm
 J/m}^2)$& \\
 \hline
 EAM&3.99&0.60&0.98&1.10&0.124&0.134&0.180&0.89\\
 \hline
 DFT&3.99&70.93&1.31&1.18&0.148&0.205&0.262&0.93\\
 \hline
Exp&4.032&1.14$^a$&1.14$^a$&1.14$^a$&0.12& & & \\
    \hline
\end{tabular}
 $^a$ Estimates for an ``average'' orientation.
\end{minipage}
\label{tab:elast} % is used to refer this table in the text
\end{table}

\subsection{Dislocation nucleation at the crack tip}

\begin{figure}[htp]
\centering
\includegraphics[width=\ww\textwidth]{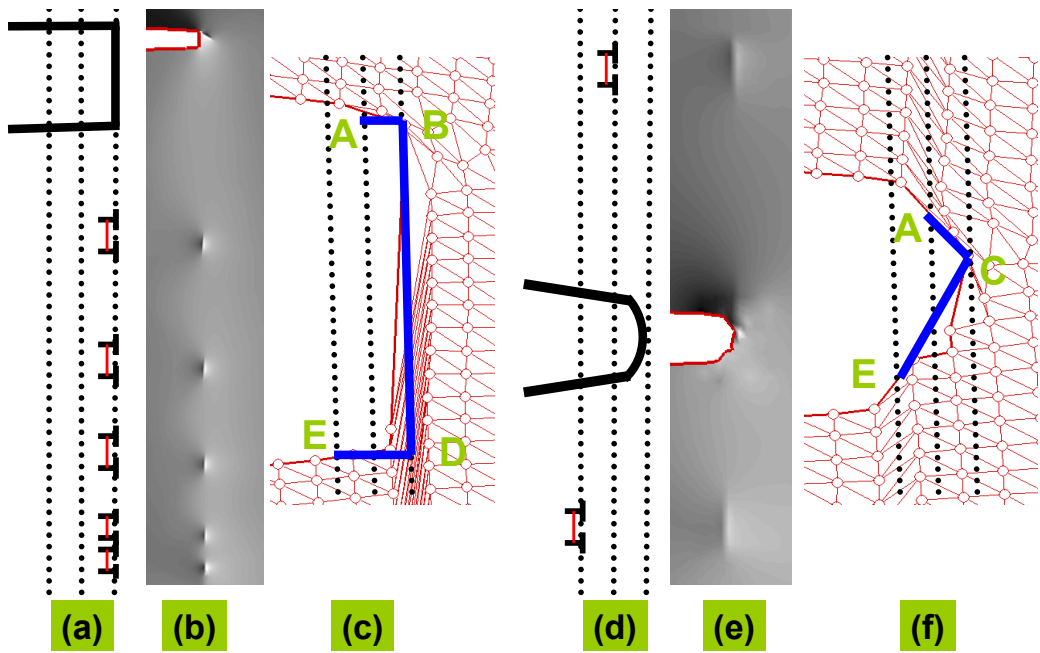}
 \caption{\label{f:comp} The schematic diagram, displacement contours and
 atomic structure with finite-elment mesh showing the dislocation pattern at the crack tip.
(a), (b) and (c) are the schematic diagram, displacement contour
plot and atomic structure respectively from EAM-QC calculation for
$K_{\rm I}=0.198 \,{\rm eV/\AA}^{2.5}$. (d), (e) and (f) are the
same from QCDFT calculation for $K_{\rm I}=0.169 \,{\rm
eV/\AA}^{2.5}$.}
\end{figure}

EAM-QC shows that the crack blunts by emitting dislocations gliding
on a single slip plane, and the number of emitted dislocations
increases as the SIF is increased. However, QCDFT predicts that two
dislocations are nucleated from the crack tip and they glide at two
adjacent slip planes. Moreover, the number of emitted dislocations
remains the same (two) for all SIFs. These distinctions are
highlighted in \fig{f:comp} where schematic diagrams, displacement
contours and atomic structures at the crack tip for the two
calculations are presented. In the schematic diagrams (a and d), the
crack tip is represented by solid black lines/curves, and the dotted
black lines denote the three relevant (111) slip planes at $x =
-3/2a, -a/2$ and $a/2$. The stacking fault between the Shockley
partials is indicated by a short red line segment. The displacement
contour plots are the same as those shown in \fig{fig:eam} and
\fig{fig:QCDFT}; they are reproduced here for convenience to the
reader. The open circles in the atomic structure plots represent
atomic positions. The sheared finite-element mesh in the atomic
structure plots is the result of passing-by dislocations (the amount
of shear corresponds to the net Burgers vector of the dislocations).
The crack tip profile is approximated by blue line segments in the
atomic structure plots. In Fig. 7(c) the crack tip profile is
approximated by rectangular line segments
$\overline{AB}$+$\overline{BD}$+$\overline{DE}$ to model the fact
that the crack tip is blunted by emitting dislocations on a single
slip plane. The active slip plane is represented $\overline{BD}$
whose position is at $x = a/2$. In Fig. 7(f), however, the crack tip
profile of QCDFT is approximated by zigzag line segments
$\overline{AC}+\overline{CE}$ because two adjacent slip planes are
activated, and their positions are at $ x = -a/2 $ and $ x=-3/2 a$.

\begin{figure}[htp]
\centering
\includegraphics[width=0.6\textwidth]{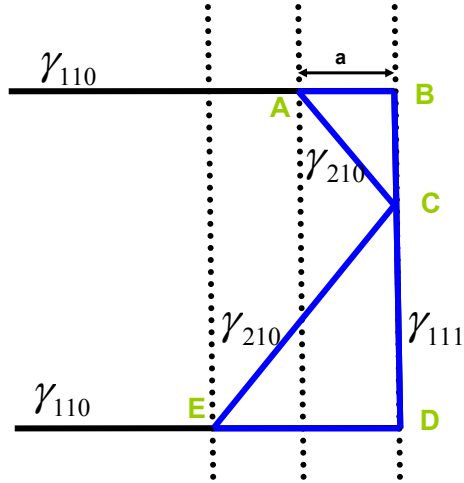}
 \caption{\label{fig:surface} A simple model for crack tip profile. The dotted line represents the relevant slip planes near the crack tip.
The rectangular line segments
$\overline{AB}$+$\overline{BD}$+$\overline{DE}$ and the zigzag
segments $\overline{AC}+\overline{CE}$ approximate EAM-QC and QCDFT
crack tip profile respectively.}
\end{figure}

To understand the origin of the differences, we resort to a simple
model that captures the essential features of the crack
configuration shown in \fig{fig:surface}. One can consult
\fig{f:comp} to understand the correspondence between the model and
the actual crack tip configuration. The reason why the rectangular
segments $\overline{AB}$+$\overline{BD}$+$\overline{DE}$ are
preferred in EAM-QC while the zigzag segments
$\overline{AC}$+$\overline{CE}$ are favored in QCDFT can be
understood from the following surface energy analysis. Without loss
of generality, we consider here the case where two dislocations are
emitted for both EAM-QC and QCDFT. The length of $\overline{BC}$
equals to the magnitude of the Burgers vector because a full
dislocation has been emitted upper-ward from A. As a result,
$\overline{AC}$ represents a \{210\} surface. There are two layers
of atoms removed in the initial crack openning, which is one Burgers
vector wide. Therefore in addition to a full dislocation emitted
downward from E, the length of $\overline{CD}$ equals to twice of
the Burgers vector magnitude. Hence $\overline{CE}$ is also a
\{210\} surface. Therefore $\overline{BD}$, $\overline{AC}$ (and
$\overline{CE}$) and $\overline{AB}$ (and $\overline{ED}$)
represents (111), \{210\} and ($\bar{1}$10) surface, respectively.
The total surface energy associated with the zigzag and rectangular
segments can be written as
$(\overline{AC}+\overline{CE})\gamma_{210}z_0$ and
$(\overline{AB}\gamma_{110}+\overline{BD}\gamma_{111}+\overline{DE}\gamma_{110})z_0$,
respectively. $z_0$ is the repeat distance along z axis. The various
surface energies have been calculated by both DFT and rescaled EAM
as summarized in Table 1. For this particular crack configuration,
the surface energy of the rectangular segments is 0.05 eV lower than
that of the zigzag segments based on EAM energetics. On the other
hand, the surface energy of the zigzag segments is 1.23 eV lower
than that of the rectangular segments according to the DFT
energetics (see note \footnote{$a=\sqrt{1/3}a_0$ where $a_0$ is the
lattice constant. $\overline{AB}=a$, $\overline{BC}=\sqrt{3/2}a$,
$\overline{BC}=\sqrt{6}a$, $\overline{DE}=2a$,
$\overline{AC}=\sqrt{5/2}a$, $\overline{CE}=\sqrt{10}a$, and
$z_0=\sqrt{9/2}a$. Taking a = 2.3036 \AA, the total surface energy
of the zigzag configuration is 3.67 eV for EAM and 3.93 eV for DFT.
The total surface energy of the rectangular segments is 3.62 eV for
EAM and 5.16 eV for DFT.}). Therefore, the zigzag segments are
preferred in QCDFT while the rectangular segments are favored in
EAM-QC. The same conclusion holds for other loadings as well.

\subsection{Deformation twinning at the crack tip}
According to the Peierls criterion of deformation twinning at a
crack tip \citep{Tadmor}, one can define twinnability which is the
likelihood of a material to twin as opposed to slip at the crack
tip. The dimensionless twinnability can be expressed as
\citep{tbtwinnability}
\begin{equation}
\tau_t =[1.136-0.151\frac{\gamma_{\rm sf}}{\gamma_{\rm us}}]
\sqrt{\frac{\gamma_{\rm us}}{\gamma_{\rm ut}}}
\end{equation}
The coefficients 1.136 and 0.151 are universal constants for an fcc
lattice. $\gamma_{\rm sf}$, $\gamma_{\rm us}$, and $\gamma_{\rm ut}$
are intrinsic stacking fault, unstable stacking fault and unstable
twinning energy respectively. A material will emit a dislocation
before twinning if $\tau_t <1$ and will twin first if $\tau_t>1$.
Our DFT and EAM calculations find that $\tau_t$ is less than 1 as
shown in Table I, which suggests that no true deformation twinning
along (111) plane be formed, consistent with the QCDFT and EAM-QC
simulation results. However, a two-layer-wide micro-twin along
($1\bar{3}\bar{1}$) plane is formed in EAM-QC while no such twin is
present in QCDFT. The distinction between the QCDFT and EAM-QC
results is due to the large discrepancy of $\gamma_{us}$, which is
the energy barrier for a leading partial nucleation at a crack tip
\citep{Rice1992}. The large DFT value of $\gamma_{us}$ renders the
nucleation of the leading partial difficult, which in turn prohibits
the formation of deformation twinning.

\begin{figure}[htp]
\centering
\includegraphics[width=\ww\textwidth]{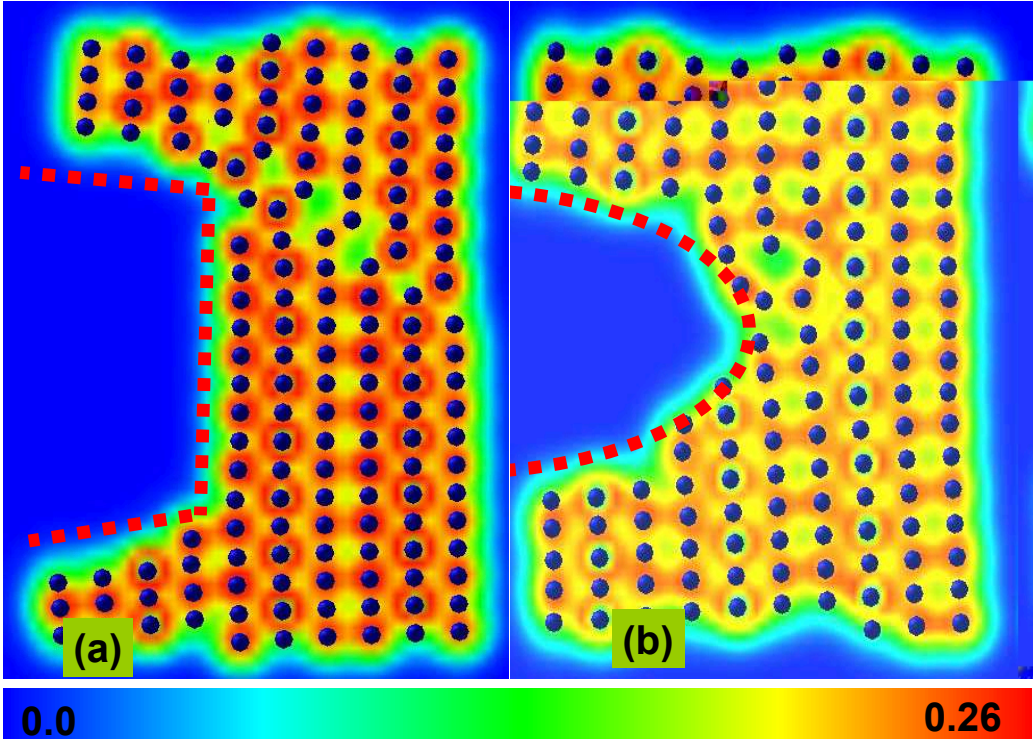}
 \caption{\label{fig:charge} The electron charge density (\AA$^{-3}$) at the crack tip for (a) EAM-QC at $K_{\rm{I}} =  0.198~ {\rm
eV/\AA}^{2.5}$ and (b) QCDFT at $K_{\rm{I}} = 0.169~ {\rm
eV/\AA}^{2.5}$. The density contours range from 0 (blue) to 0.26
(red). The blue sphere stands for atomic position and dashed line is
an approximate profile of the charge density contours.}
\end{figure}

\subsection{Electron density at the crack tip}
In Fig. 10, we present the electron charge density around the crack
tip for the EAM-QC configuration of $K_{\rm{I}} =  0.198~ {\rm
eV/\AA}^{2.5}$ (left) and the QCDFT configuration of $K_{\rm{I}} =
0.169~ {\rm eV/\AA}^{2.5}$ (right). The charge density for EAM-QC is
determined by a superposition of atomic densities (obtained by VASP)
centered at each EAM atoms. The distortion of charge density due to
the defects is clearly visible. In EAM-QC the atomic bonding is
weakened along the twin plane while in QCDFT the atomic bonding is
significantly disrupted at the center of the crack. More
importantly, the charge density profile of QCDFT is smoother than
that of EAM-QC as indicated by the dashed curves. The ``sharp"
corners of EAM-QC charge density lead to higher kinetic energy of
electrons. Since EAM-QC does not involve quantum mechanics, the
``sharp corner"  is not penalized energetically and thus
permissible. Of course, the electron charge density profile reflects
the underlying atomic structure: ``sharp corners" correspond to a
straight crack front thanks to a single active slip plane; smooth
corners correspond to a more rounded crack front.

\section{CONCLUSION}
In summary, we have carried out a comparative study of fracture in
Al by using two distinctive atomic interactions: quantum mechanical
density functional theory and empirical embedded atom method. The
DFT description of the crack tip is achieved by QCDFT method while
the empirical description by EAM-QC method. In addition to
quantitative differences, qualitatively different fracture behavior
is also observed between the two methods. EAM-QC predicts a more or
less rectangular crack tip configuration while QCDFT yields a more
rounded tip profile. The difference is due to the fact that the
emitted dislocations glide on a single slip plane in EAM-QC while
two adjacent slip planes are active in QCDFT. As the stress
intensity factor is increased, more and more dislocations are
emitted from the crack tip in EAM-QC while the number of
dislocations remains the same up to the maximum loading applied in
QCDFT calculations. A micro-twin is observed at the crack tip in
EAM-QC, but it is absent in QCDFT. The electron density profile at
the crack tip is also different between EAM-QC and QCDFT. All these
differences can be understood in terms of defect energetics,
including surface energy and stacking fault energy.

The different results received suggest that the atomic nature of a
crack tip is important and an accurate description of the atomic
interaction at the crack tip is indispensable. Although empirical
potentials can be developed by fitting to DFT results, it is
unlikely they will reproduce all {\it relevant} DFT energetics. This
is particularly so since one does not know {\it a priori} what are
the relevant energetics for a given crack. If several chemical
species are present in a crack tip, the task of fitting a
satisfactory potential becomes even more daunting. Therefore the
solution lies at an explicit quantum mechanical description of the
crack tip, most likely in a form of DFT-based multiscale modeling,
such as QCDFT. The present paper concerns atomistic aspect of
fracture which is important for many purposes. However, there are
interesting fracture phenomena that do not depend on atomistic
features and thus are beyond the scope of the present paper.
Finally, we have not touched upon fracture dynamics. The questions -
such as will finite temperature dynamics amplify or diminish the
differences that we observed and what are the best strategies to
implement dynamics in a multiscale setting - remain unanswered. We
hope that the present paper could spawn more research effort in
answering these questions.

%\begin{acknowledgments}
\section{ACKNOWLEDGEMENTS} We thank Ellad Tadmor for his assistance
of constructing the crack model and many helpful discussions. The
work at California State University Northridge was supported by NSF
PREM grant DMR-0611562 and DoE SciDAC grant DE-FC02-06ER25791.
%\end{acknowledgments}

\bibliographystyle{elsarticle-harv}
\bibliography{crack09}

\end{document}